\documentclass[11pt]{article}
\usepackage{amssymb}
\usepackage{epsfig}
\renewcommand \baselinestretch{1.3}
\renewcommand{\thesection}{\arabic{section}.}
\newcommand{\half}{{\scriptstyle{\frac{1}{2}}}}
\newcommand{\LP}{\lambda \Phi^4}
\newcommand{\dl}{\delta^{(3)}({\bf r})}
\newcommand{\del}{{\mbox{\boldmath $\nabla$}}}
\newcommand{\BE}{\begin{equation}}
\newcommand{\EE}{\end{equation}}
\newcommand{\BA}{\begin{eqnarray}}
\newcommand{\EA}{\end{eqnarray}}
\newcommand{\vol}{{\sf V}}
\newcommand{\num}{{\sf N}}
\newcommand{\boldsymbol}{\bf}

\begin{document}
\begin{titlepage}

\vspace*{1mm}
\begin{center}

            {\LARGE{\bf Old and new ether-drift experiments: \\
a sharp test for a preferred frame }}

\vspace*{14mm}
{\Large  M. Consoli and E. Costanzo}
\vspace*{4mm}\\
{\large
Istituto Nazionale di Fisica Nucleare, Sezione di Catania \\
Dipartimento di Fisica e Astronomia dell' Universit\`a di Catania \\
Via Santa Sofia 64, 95123 Catania, Italy}
\end{center}
\begin{center}
{\bf Abstract}
\end{center}
Motivated by the critical remarks of several
authors, we have re-analyzed the classical ether-drift
experiments with the
conclusion that the small observed deviations should not be neglected. 
In fact, within the framework of Lorentzian Relativity, they might indicate
the existence of a preferred frame relatively to which 
the Earth is moving with a 
velocity $v_{\rm earth}\sim 200$ km/s (value 
projected in the
plane of the interferometer). We have checked this idea by comparing
with the
modern ether-drift experiments, those where the observation of the fringe
shifts is replaced by the difference $\Delta \nu$
in the relative frequencies of two cavity-stabilized lasers, upon local
rotations of the apparatus or under the Earth's rotation. It turns out that, 
even in this case, 
the most recent data are consistent with the same value of the
Earth's velocity, once the vacuum within the cavities is considered a physical
medium whose refractive index is fixed by General Relativity.
 We thus propose a sharp experimental test that can 
definitely resolve the issue. If the small deviations observed in the classical
ether-drift experiments were not mere instrumental artifacts, by
replacing the high vacuum in the resonating cavities 
with a dielectric gaseous medium ({\it e.g.} air), the typical measured 
$\Delta\nu\sim 1$ Hz should increase {\it by orders of magnitude}. This 
expectation is consistent with the characteristic modulation of a few kHz
observed in the original experiment with He-Ne masers. 
However, if such enhancement would not be confirmed by new and more
precise data, the existence of a preferred frame can be definitely ruled out.
 
\end{titlepage}

\section{Introduction}

There are two basically different interpretations of the
Theory of Relativity. On one hand, there is Einstein's Special Relativity
\cite{einstein}. 
On the other hand, there is the `Lorentzian' approach where, following the
original Lorentz
and Poincar\`e point of view \cite{lorentz,poincare}, the same
relativistic effects between two observers, rather than being due to their
relative motion, might be interpreted in terms of
their {\it individual} motion with respect to a preferred
frame.

Today the former interpretation is generally
accepted. However, the potential consequences of retaining
a physical substratum as an important element of the physical theory 
\cite{martin}, may induce to re-discover the
implications of the latter.
For instance, replacing the empty space-time of
Special Relativity with a preferred frame, one gets
a different view of the non local aspects of the quantum theory, 
see Refs.\cite{hardy,scarani}.

Another argument that might induce to re-consider
the idea of a preferred frame was given in ref.\cite{pagano}. The argument
was based on the simultaneous presence of two ingredients that are often
found in present-day elementary particle physics, namely: a) vacuum 
condensation, as with the Higgs field in the electroweak theory, and b) an
approximate form of locality, as with cutoff-dependent, effective quantum
field theories. In this case, one is faced with `reentrant violations of
special relativity in the low-energy corner' \cite{volo}. These are deviations
at {\it small} momenta $|{\bf{p}}| < \delta$ where the infrared scale $\delta$
vanishes, in units of the Lorentz-invariant scale $M$ of the theory, 
only in the local limit of the continuum theory
${{\Lambda}\over{M}} \to \infty$, $\Lambda$ being the
ultraviolet cutoff. A simple interpretation of the phenomenon, 
 in the case of a condensate
of spinless quanta, is in terms of density fluctuations of the 
system \cite{stevenson,consoli}, the continuum theory corresponding to the
incompressibility limit. The resulting
picture of the ground state is closer to a medium with a non-trivial
refractive index \cite{pagano} than to the empty space-time of
Special Relativity. Therefore, in the presence of a non-trivial vacuum, it
is perfectly legitimate to ask whether the {\it physically realized} form
of the Theory of Relativity is closer to the Einstein's formulation or to
the original point of view with a preferred frame and try to get the answer 
from experiments. 

For a modern presentation of the Lorentzian approach, one can follow Bell
\cite{bell,brown} and introduce a preferred reference frame
$\Sigma$, with coordinates $(X,Y,Z,T)$, for which time is homogeneous and
space is homogeneous and isotropical. $\Sigma$ is a preferred frame since
the relative motion with respect to it introduces 
{\it physical} modifications of all length and time measuring devices.  
This means, for instance, that when atoms are (`gently') set in motion 
their basic parameters are modified by the Larmor time-dilation factor and 
by the Fitzgerald-Lorentz length contraction along the direction of 
motion. 

One can introduce, however, a primed set of variables $(x',y',z',t')$
in terms of which the description of the moving atoms coincides with that of 
the stationary atoms in terms of the original $(X,Y,Z,T)$ coordinates. 
The transformation from $(X,Y,Z,T)$ to $(x',y',z',t')$ is precisely the
standard Lorentz transformation in terms, say, of a dimensionless velocity
parameter 
$\beta'=v'/c$ (we restrict for simplicity to one-dimensional motions). 
In this way, the homogeneity and isotropy of space-time
hold for $S'$ as well. 

Now, since Lorentz transformations have a group
structure, the relation between two observers $S'$ and $S''$,
associated respectively with coordinates $(x',y',z',t')$ and
$(x'',y'',z'',t'')$ and individual velocity parameters $\beta'$ and $\beta''$, 
is also a Lorentz transformation with relative velocity parameter 
$\beta_{\rm rel}$ given by
\BE
          \beta_{\rm rel}= {{\beta' - \beta''}\over{ 1- \beta' \beta''}}
\EE
Therefore, the crucial question to test the existence of a preferred frame
is the following: can the individual parameters 
$\beta'$ and $\beta''$ be determined separately through ether-drift 
experiments ?
The standard `null-result' interpretation of the Michelson-Morley \cite{mm}
experiment means that this is not possible. 
Therefore, if really only $\beta_{\rm rel}$ is experimentally measurable, 
one is driven to conclude (as Einstein did in 1905 
\cite{einstein}) that the introduction of a preferred frame is `superfluous',
all effects of $\Sigma$ being re-absorbed into the relative space-time
units of any pair $(S',S'')$. 

On the other hand, if the Michelson-Morley experiment would give a non-null
result, so that $\beta'$ and $\beta''$ can be separately determined, then
the situation is completely
different. In fact, now $\beta_{\rm rel}$ is a derived quantity
and the Lorentzian point of view is uniquely singled out. This possibility
should be considered seriously since Einstein, in his 1905 article
\cite{einstein}, was explicitely referring to ``...the unsuccessful attempts 
to discover any motion of the earth relatively to the light medium''. Since
a physical theory is not just an axiomatic structure but is founded on some
basic experimental facts, it is obvious that Einstein would have argued 
differently knowing that the Michelson-Morley data actually give a 
non-null result. 

The aim of this paper is to critically re-analyze the classical and 
modern ether-drift experiments starting from the original Michelson-Morley 
experiment. Our main motivation is that, 
according to some authors, the null-result interpretation of that
experiment is not so obvious. 
The observed fringe shifts, while
certainly smaller than the classical prediction corresponding to the orbital
velocity of the Earth, were {\it not} negligibly small. 
This point was clearly expressed by Hicks \cite{hicks}
and also by Miller, see fig.4 of ref.\cite{miller}. In the latter case, 
Miller's refined analysis of the half-period, 
second-harmonic effect observed in the original experiment, 
and in the subsequent ones by Morley and Miller \cite{morley}, 
showed that all data were consistent with an
effective, {\it observable} velocity lying in the range 7-10 km/s. 
For comparison, the Michelson-Morley experiment gave a value
$v_{\rm obs} \sim 8.8 $ km/s for the noon observations 
and a value $v_{\rm obs} \sim 8.0 $ km/s for the evening observations.

Now, since these velocities are non-zero, an
interpretation of the experiment requires
a theoretical framework to relate
the `kinematical' Earth's velocity $v_{\rm earth}$, relatively to 
$\Sigma$, to its observable value $v_{\rm obs}$ effectively governing
the magnitude of the fringe shifts seen in the interferometer. 
In this sense, the
classical pre-relativistic prediction 
$v_{\rm obs}=v_{\rm earth}$ 
might be dramatically wrong and, with it, 
the assumed null-result interpretation of the experiment. 

Motivated by the previous remarks, we have first re-considered in this paper
the Michelson-Morley 
original data and re-calculated the values of $v_{\rm obs}$ for their 
experiment. Our findings completely confirm Miller's indications of
an average observable velocity $v_{\rm obs}\sim 8.4$ km/s. 

Further assuming, as in the pre-relativistic physics, 
the existence of a preferred reference frame $\Sigma$ where light propagates 
isotropically, but correctly
using Lorentz transformations (instead of Galilei's transformations) 
to connect $\Sigma$ to the Earth's reference frame, 
it turns out that this $v_{\rm obs}$
corresponds to a {\it real} Earth's velocity, 
in the plane of the interferometer, 
$v_{\rm earth} \sim 200$ km/s. 

We emphasize that the use of Lorentz transformations is absolutely crucial. 
In fact, in this case, differently from the classical prediction
$v_{\rm obs}=v_{\rm earth}$, the fringe
shifts measured with an interferometer operating in a dielectric medium 
of refractive index ${\cal N}_{\rm medium}$ are proportional to the Fresnel's
drag coefficient $1- 1/{\cal N}^2_{\rm medium}$. Therefore, a rather
large `kinematical' velocity $v_{\rm earth} \sim 200$ km/s
is seen, in an in-air-operating optical
system, as a small `observable' velocity $v_{\rm obs}\sim 8.4$ km/s. 
At the same time, the same $v_{\rm earth} \sim 200$ km/s
becomes an effective $v_{\rm obs}\sim 3$ km/s for the Kennedy's and
Illingworth experiments (performed in an apparatus filled with helium) or
a $v_{\rm obs}\sim 1$ km/s 
for the Joos experiment (performed in an evacuated housing), in agreement
with the experimental results. 
Finally, such a value $v_{\rm earth} \sim 200$ km/s, 
 deduced from the {\it absolute magnitude}
of the fringe shifts, is also
consistent with the typical range of kinematical velocities 
195 km/s $\leq v_{\rm earth} \leq $ 211 km/s
(see table V of ref.\cite{miller}) needed by Miller 
to describe the {\it variations} of the
ether-drift effect in different epochs of the year. 

After this first part, we have concentrated our analysis on 
the modern ether-drift experiments, those 
where the observation of the interference
fringes is replaced by the difference $\Delta \nu$
in the relative frequencies of two cavity-stabilized lasers upon local
rotations of
the apparatus \cite{brillet} or under the Earth's rotation \cite{muller}. It
turns out that, even in this case, 
the most recent data \cite{muller} leave some space for a non-null 
interpretation of the experimental results with the same Earth's velocity
(in the plane of the interferometer) 
$v_{\rm earth} \sim 200$ km/s extracted
from the classical experiments. 

For this reason, and as conclusion of our analysis, we shall  
propose a sharp experimental test that can definitively decide about
the existence of a preferred frame. If the small deviations found in the
classical experiments were not mere instrumental artifacts, 
by replacing the high vacuum used in the 
resonating cavities with a dielectric gaseous medium, the typical
frequency of the signal should increase 
from values $\Delta\nu\sim 1$ Hz up to $\Delta\nu \sim 100$ kHz, using air, or
up to $\Delta\nu \sim 10$ kHz, using helium. The latter 
prediction appears to be consistent with the characteristic modulation of a 
few kHz in the magnitude of the $\Delta\nu$'s 
observed by Jaseja et {\it al.} \cite{jaseja} using He-Ne masers. 

The plane of the paper is as follows. In Sect.2 we shall present 
our re-analysis of the Michelson-Morley original data. 
In Sect.3 we shall illustrate the role of Lorentz transformations and, in 
Sect.4, discusss how they can be used consistently to 
connect the Michelson-Morley, Morley-Miller and Miller's experiments to
those performed by Kennedy, Illingworth and Joos. 
Later, in Sect.5 we shall address the present-day
experiments and, finally, in Sect.6 present our conclusions. 

\section{The Michelson-Morley data}

For the importance of the issue and to provide the reader with all essential
ingredients of the analysis, we have re-considered 
the original data obtained by Michelson and Morley in each of
the six different sessions of their experiment. 
No form of inter-session averaging has been attempted. 
Following this procedure, there are sizeable differences with 
       respect to the original analysis of Michelson-Morley
\cite{mm} or with respect to the more recent paper by Handschy \cite{handschy}. 
  The reason was pointed out by Hicks \cite{hicks} long time ago:
     one is not allowed to average data of different sessions unless 
     one is sure that the direction of the ether-drift effect remains the same
     (see page 34 of \cite{hicks} ``It follows that averaging the results of
     different days in the usual manner is not allowable...If this is not 
     attended to, the average displacement may be expected to come out 
     zero...''). 

In other words, the ether-drift, if it exists, has
a vectorial nature. Therefore, rather than averaging
the raw data from the various sessions, one should first consider
the data from the i-th experimental session and extract the observable velocity 
$v_{\rm obs}(i)$ and the ether-drift direction $\phi_2(i)$ 
for that session. Finally, a mean magnitude 
$\langle v_{\rm obs}\rangle$ and a mean direction $\langle \phi_2 \rangle$
can be obtained by averaging the individual determinations
(see figs. 22 of ref.\cite{miller}).

Now, when the raw data of different sessions are not averaged, 
       the observable velocity comes out to be larger, its error becomes 
       smaller so that the evidence for an ether-drift effect becomes stronger
(see page 36 of  ref.\cite{hicks} `` ...this naturally leads 
     to the reconsideration of the numerical data obtained by Michelson 
     and Morley, who did lump together the observations taken in 
     different days. I propose to show that, instead of giving a null 
     result, the numerical data published in their paper show distinct 
     evidence of an effect of the kind to be expected''). 

After Hicks, the same conclusion was drawn by Miller.
For instance, in the Morley-Miller data \cite{morley}, 
the morning and evening observations each were indicating
 an effective velocity of
about 7.5 km/s (see fig.11 of ref.\cite{miller}). This indication
 was completely lost after 
averaging the raw data as in ref.\cite{morley}. Finally, 
the same point of view has been advocated by Munera in his recent 
re-analysis of the classical experiments \cite{munera}. 

To obtain the fringe shifts, we have
followed the well defined procedure adopted in the classical experiments as
described in Miller's paper \cite{miller}. Namely, starting from the seventeen
entries, say $E(i)$, reported in the table of ref.\cite{mm}, 
one was first correcting the data for the large linear drift responsible for
the difference $E(1) -E(17)$ between the 1st entry and the 17th entry
obtained after a complete rotation of the apparatus.
In this way, one was adding 15/16 of the correction
to the 16th entry, 14/16 to the 15th entry and so on, 
thus obtaining a set of 16 corrected entries 
\BE
E_{\rm corr}(i)={{i-1}\over{16}} (E(1)-E(17)) + E(i)
\EE
Finally, the fringe shift is defined from the 
differences between each of the corrected entries $E_{\rm corr}(i)$ and 
their average value $\langle E_{\rm corr} \rangle$ as
\BE
 {{\Delta \lambda (i)}\over{\lambda}}= 
E_{\rm corr}(i) - \langle E_{\rm corr} \rangle
\EE
These final data for each session are reported in table 1.

With this procedure, 
the fringe shifts are given as a periodic function (with vanishing mean) 
in the
range $0 \leq \theta \leq 2\pi$, with $\theta={{i-1}\over{16}} 2\pi$, 
so that they can be reproduced in a Fourier expansion
\BE
\label{fourier}
      {{\Delta \lambda(\theta)}\over{\lambda}} =\sum_n~{A}_n 
\cos(n\theta- n\phi_n)
\EE
The Fourier analysis allows to 
determine the direction (`azimuth') of the ether-drift effect, 
from the phase $\phi_2$ of the second-harmonic component, and an observable velocity 
from the value of its amplitude (see for instance the classical analysis of
Refs.\cite{hicks,kennedy}).
To this end, we have used the basic relation of the experiment
\BE
\label{a2class}
                 2 {A}_2= 
{{2D}\over{\lambda}} ~{{v^2_{\rm obs} }\over{c^2}}
\EE
where $D$ is the length of each arm of the interferometer. 
In the classical theory (see for instance Refs.\cite{hicks,kennedy}), where
the space-time transformations connecting the Earth's frame 
to the preferred frame are Galilei's transformations, 
the observable velocity $v_{\rm obs}$ coincides with the kinematical Earth's 
velocity $v_{\rm earth}$ (value projected in the plane of the interferometer).

Notice that, as emphasized by Shankland et {\it al.} (see page 178 of 
ref.\cite{shankland}), it is the quantity 
$2 {A}_2$, and not ${A}_2$ itself, 
that should be compared with the maximal
displacement obtained for rotations of the apparatus through 
$90^o$ in its optical plane (see also eq.(\ref{fringe}) 
below). Notice also that the quantity $2{A}_2$ is denoted by $d$ in 
Miller's paper (see page 227 of ref.\cite{miller}). 

As pointed out by Hicks long ago, 
there is a theoretical motivation for a large full-period, first-harmonic effect 
in the experimental data. Its theoretical interpretation is
in terms of the `actual' (as opposed to `ideal') 
    arrangements of the mirrors \cite{hicks}. 
As such, this effect is not a form of
    `background' but has to be present in the outcome of real 
    experiments. For more details, see the 
discussion given by Miller, in particular 
    fig.30 of ref.\cite{miller}, where it is shown that his observations were 
    well consistent with Hicks' theoretical study. The observed
first-harmonic effect is sizeable, 
of comparable magnitude or even larger than the 
    second-harmonic effect. 
The same conclusion was also obtained 
    by Shankland et {\it al.} in their re-analysis of the 
Miller's data.

We have reported in table 2, our values of
       ${A}_2$ for each session. The individual determinations, that
show good consistency, have been obtained from a
10-parameter fit to the various sets of 16 data (see fig.1) 
where, following Miller's indications, the first five harmonics were included. 
To test the stability of these ${A}_2$ values, we have also
fitted the even combination of fringe shifts
  ${{\Delta \lambda(\theta) + \Delta \lambda(\pi+\theta)}\over{2\lambda}} $. 
In this second type of fit, where 
only the even harmonics appear, the central values of
${A}_2$ come out exactly as in table 2 with slightly smaller
errors ($\pm 0.004$ rather than $\pm 0.005$)
and the fourth-harmonic component is consistent
with the background (see fig.2). 

While the individual values of $A_2$ show good consistency, 
there are large fluctuations in the values of $\phi_2$ for the various
sessions. 
 Their typical trend is in qualitative agreement with the values
reported by Miller. For this comparison, see fig.22 of ref.\cite{miller}, in
particular the large scatter of the data taken around
August 1st, as this represents the epoch of the year which is closest 
to the period of July when 
the Michelson-Morley observations were actually performed. Just
this type of `erratic' behaviour motivated Miller's idea that a very 
large number of measurements, performed during the whole 24 hours of the day, 
was needed for a reliable
determination of the azimuth of the ether-drift effect. 

Concerning the extraction of the observable velocity, we note that
for the Michelson-Morley apparatus where
 ${{D}\over{\lambda}}\sim 2\cdot 10^7$ \cite{mm}, 
 it becomes convenient to normalize the experimental values of
${A}_2$ to the classical prediction for an Earth's velocity of 30 
km/s 
\BE
\label{classical}
{{D}\over{\lambda}} ~{{(30 {\rm km/s})^2 }\over{c^2}}\sim 0.2
\EE
and we obtain
\BE
\label{aa2}
v_{\rm obs}  \sim 30 ~ \sqrt { 
{{A}_2 }\over{ 0.2 } }~~{\rm km/s} 
\EE
Now, by inspection of table 2, we find that
the average value of ${A}_2$ from the noon sessions, 
${A}_2=0.017 \pm 0.003$, indicates a velocity
$v_{\rm obs}=8.7 \pm 0.8$ km/s
and the average value from the evening sessions, ${A}_2=0.014 \pm 0.003$, 
indicates a velocity $v_{\rm obs}=8.0 \pm 0.8$ km/s. Since 
the two determinations
are well consistent with each other, we conclude that the 
Michelson-Morley experiment provides an average
${ A}_2$ which is $\sim 1/13$ of the
classical expectation and an average observable velocity
       $v_{\rm obs} \sim 8.4 \pm 0.5~{\rm km/s}$ in excellent agreement 
with Miller's analysis of the Michelson-Morley data.

On the other hand, the comparison with the interpretation that Michelson and 
Morley gave of their data is not so simple. First of all, they 
averaged the raw data from the various sessions 
so that the evidence for an ether-drift are unavoidably
weaker. In addition, they
do not quote any mean velocity but just start from the observation that
``...the displacement to be expected was 0.4 fringe'' while ``...the
actual displacement was certainly less than the twentieth part of this''.
In this way, 
since the displacement is proportional to the square of the velocity,
``...the relative velocity of the earth and the ether is...
certainly less than one-fourth of the orbital earth's velocity''.

The straightforward translation of this upper bound is
$v_{\rm obs} < $ 7.5 km/s.
However, even accepting their average of the raw data, 
their estimate is likely affected by a theoretical uncertainty. 
In fact, in their fig.6, Michelson and Morley reported their experimental 
fringe shifts together with the plot of a reference
second-harmonic component. In doing so, they plotted a wave 
with amplitude 
$A_2=0.05$, that they interpret as {\it one-eight} of the theoretical
displacement expected on the base of classical physics, 
thus implicitely assuming $A^{\rm class}_2$=0.4. 
As shown in eq.(\ref{classical}), 
the amplitude of the classically
expected second-harmonic component is {\it not} 0.4
but is just one-half of that, {\it i.e.} 0.2. Therefore, 
their experimental upper bound
$A^{\rm exp}_2 < {{0.4}\over{20}}=$0.02, using our eq.(\ref{aa2}), might also
be interpreted as
$v_{\rm obs} <$ 9.5 km/s, consistently with our estimate
       $v_{\rm obs} \sim 8.4 \pm 0.5~{\rm km/s}$.

We conclude this section noticing that our Michelson-Morley value 
       $v_{\rm obs} \sim 8.4 \pm 0.5~{\rm km/s}$ is also
in good 
agreement with the experimental results obtained by 
Miller {\it himself} at Mt. Wilson. As anticipated, 
 differently from the original Michelson-Morley experiment, 
Miller's data were taken over the 
entire day and in four epochs of the year. 
However, after the critical re-analysis of Shankland et {\it al.} \cite{shankland}, 
it turns out that
the average daily determinations of ${A}_2$ for the four epochs
were statistically consistent (see page 170 of ref.\cite{shankland}).
Therefore, one can
average the four daily determinations, ${A}_2=0.044 \pm 0.005$, 
and compare with the equivalent form of eq.(\ref{classical})
for the Miller's interferometer
${{D}\over{\lambda}} ~{{(30 {\rm km/s})^2 }\over{c^2}}\sim 0.56$. Again, 
the observed ${A}_2$ is $\sim 1/13$ of 
the classical expectation for an Earth's velocity of 30 km/s and 
the effective 
$v_{\rm obs}$ is {\it exactly} 
the same as for the Michelson-Morley
data. 

This close agreement 
is also confirmed by another independent analysis. In fact, M\'unera's analysis
\cite{munera}
of the only Miller's set of data explicitely reported in the literature 
yields the value 
$v_{\rm obs}=8.2 \pm 1.4$ km/s (errors at the 95$\%$ C.L.), 
again in excellent agreement with our value for the Michelson-Morley
experiment.

Therefore, 
by also taking into account the results obtained by Morley-Miller in the years
1902-1905, 
as shown in fig.4 of ref.\cite{miller}, we conclude that the
results of these three main classical 
ether-drift experiments can be summarized into the value
\BE
\label{vobs}
       v_{\rm obs} \sim 8.5 \pm 1.5~{\rm km/s}
\EE

\section{The role of Lorentz transformations}

Now, suppose we accept 
the value in eq.(\ref{vobs}) to summarize the results of the
Michelson-Morley, Morley-Miller and Miller experiments. 
As these were performed in air, it would mean that 
the measured 
two-way speed of light differs from an exactly isotropical value
\BE
\label{air}
u_{\rm air}={{c}\over{\cal{N}_{\rm air} }}
\EE
${\cal N}_{\rm air}$ denoting the refractive index of the air. Namely, 
for an observer placed on the Earth light is slightly anisotropical 
at a level
${\cal O}({{v^2_{\rm obs}}\over{c^2}}) \sim 10^{-9}$ so that
eq.(\ref{air}) is only accurate at a lower level of accuracy, say
$\sim 10^{-8}$. 

On the other hand, for the Kennedy's \cite{kennedy0} experiment, 
where the whole
optical system was inclosed in a sealed metal case containing helium at 
atmospheric pressure, the observed anisotropy was definitely smaller. In fact, 
the accuracy of the experiment, such to exclude
 fringe shifts as large as 1/4 of those expected 
on the base of eq.(\ref{vobs}) (or 1/50 of that expected on the base of a
velocity of 30 km/s) allows to place an upper bound
$v_{\rm obs} < 4$ km/s. This is confirmed by the re-analysis of the
Illingworth's experiment \cite{illing} performed by M\'unera \cite{munera}
who pointed out some incorrect assumptions in the original analysis of the 
data. From this re-analysis, the relevant observable velocity turns out to be 
$v_{\rm obs}= 3.1 \pm 1.0$ km/s (errors at the 95$\%$ C.L.) 
\cite{munera}, with typical fringe shifts that were 1/100 of that expected
for a velocity of 30 km/s.
 Again, this means that, 
for an apparatus filled with gaseous helium at atmospheric pressure, 
the measured two-way speed of light differs from the exactly isotropical value
${{c}\over{\cal{N}_{\rm helium} }}$ by terms
${\cal O}({{v^2_{\rm obs}}\over{c^2}}) \sim 10^{-10}$. 

Finally, for the Joos experiment \cite{joos}, performed in an evacuated
housing and where any ether-wind was found smaller than $1.5$ km/s, 
the typical value $v_{\rm obs}\sim 1$ km/s means
that, in that particular type of vacuum, the fringe shifts were smaller than
1/400 of those expected for an Earth's velocity of 30 km/s and
the anisotropy of the two-way speed of light was at the 
level $\sim 10^{-11}$. 

Tentatively, we shall try to summarize the above experimental 
results saying that when light propagates in 
a gaseous medium, the exactly isotropical value
\BE
\label{u}
u={{c}\over{\cal{N}_{\rm medium} }}
\EE
holds approximately for an observer placed on the Earth. Apparently, 
the observed trend is such
that the anisotropy becomes smaller when the refractive index of the medium 
approaches unity. In fact $v_{\rm obs}$, and thus the anisotropy,
is larger for those interferometers operating in air, where
${\cal N}_{\rm air} \sim 1.00029$, and becomes smaller in 
experiments performed in helium, where
${\cal N}_{\rm helium}\sim 1.000036$, or in an evacuated housing. 
This observation suggests to interpret the experiments 
adopting the point of view of ref.\cite{pagano} that
we shall briefly recapitulate in the following.

A small
anisotropy of the two-way speed of light measured by an observer $S'$ placed
on the Earth, leads to consider, as in the pre-relativistic physics, 
the existence of a preferred reference frame $\Sigma$, where light propagates
isotropically, and generate the anisotropy in $S'$ as a consequence of the
relative motion. This is similar to 
the conventional treatment of the Michelson-Morley experiment where
one starts from the isotropical value $c$ in 
$\Sigma$ and uses Galileian relativity
(for which the speed of light becomes $c\pm v$)
to transform to the observer $S'$ placed in the Earth's frame. 

In doing so, however, 
one neglects i) that light may propagate in 
a dielectric medium and ii) that Galilei's trasformations have to be
replaced by Lorentz transformations. 
These preserve the value of the speed of light 
{\it in the vacuum} $c=2.9979...\cdot10^{10}$ cm/s
but do not preserve its isotropical value 
in a medium. In this case, one has to account for 
a non-vanishing Fresnel's drag coefficient
\BE
 k_{\rm medium}=
1- {{1}\over{ {\cal N}^2_{\rm medium} }} \ll 1
\EE
Therefore, to generate an anisotropy in $S'$ one can start
from eq.(\ref{u}), assumed to be valid in $\Sigma$, and
apply a Lorentz transformation. By denoting 
${\bf{v}}$ the velocity of
$S'$ with respect to $\Sigma$, the general Lorentz transformation 
that gives the one-way speed of light in $S'$ is
($\gamma= 1/\sqrt{ 1- {{ {\bf{v}}^2}\over{c^2}} }$) 
\BE
\label{uprime}
  {\bf{u}}'= {{  {\bf{u}} - \gamma {\bf{v}} + {\bf{v}}
(\gamma -1) {{ {\bf{v}}\cdot {\bf{u}} }\over{v^2}} }\over{ 
\gamma (1- {{ {\bf{v}}\cdot {\bf{u}} }\over{c^2}} ) }}
\EE
where $v=|{\bf{v}}|$. By keeping terms up 
to second order in $v/u$, denoting by
$\theta$ the angle between ${\bf{v}}$ and ${\bf{u}}$
 and defining $u'(\theta)= |{\bf{u'}}|$, 
we obtain
\BE
  {{ u'(\theta) }\over{u}}= 1- \alpha {{v}\over{u}} -\beta {{v^2}\over{u^2}}
\EE
where 
\BE
   \alpha = k_{\rm medium} \cos \theta + 
{\cal O} (k^2_{\rm medium} )
\EE
\BE
\beta = k_{\rm medium} P_2(\cos \theta) +
{\cal O} ( k^2_{\rm medium} )
\EE
with $P_2(\cos \theta) = {{1}\over{2}} (3 \cos^2\theta -1)$.

Finally, the two-way speed of light is 
\BE
\label{twoway}
{{\bar{u}'(\theta)}\over{u}}= {{1}\over{u}}~ {{ 2  u'(\theta) u'(\pi + \theta) }\over{ 
u'(\theta) + u'(\pi + \theta) }}= 1- {{v^2}\over{c^2}} ( A + B \sin^2\theta) 
\EE
where 
\BE 
\label{ath}
   A= k_{\rm medium}  + {\cal O} ( k^2_{\rm medium} )
\EE
and 
\BE
\label{BTH}
     B= -{{3}\over{2}} k_{\rm medium} 
+ {\cal O} (k^2_{\rm medium} )
\EE
In this way, as shown in ref.\cite{pagano}, one obtains formally the same
pre-relativistic expressions where the
kinematical velocity $v$ is replaced by an effective observable velocity
\BE
\label{vobs0}
           v_{\rm obs}= v
\sqrt { k_{\rm medium} }
 \sqrt{3} \sim v \sqrt{ -2B}
\EE
For instance, 
for the Michelson-Morley experiment, and for an ether wind
along the $x$ axis, 
the $S'$-prediction for the fringe shifts at a given angle 
$\theta$ with the $x$ axis
has the particularly simple form ($D$ being the length for $S'$ of each arm
of the interferometer) 
\BE 
\label{fringe}
{{\Delta \lambda (\theta)}\over{\lambda}}=
{{u}\over{\lambda}} (
{{2D }\over{\bar{u'}(\theta)}}- {{2 D }\over{\bar {u'}(\pi/2+\theta)}})\sim 
 {{ D }\over{\lambda}} {{v^2}\over{c^2}} (-2B) \cos (2\theta) =
 {{ D }\over{\lambda}} {{v^2_{\rm obs}}\over{c^2}}\cos (2\theta) 
\EE
that corresponds to a pure second-harmonic 
effect as in eq.(\ref{a2class})
where $v^2$ is replaced by $v^2_{\rm obs}$. 
Notice that, as discussed in the Introduction, 
in agreement with the basic isotropy of space, the measured
length of an interferometer at rest in $S'$ is $D$ 
regardless of the angle $\theta$ of its orientation. 

We observe that eqs.(\ref{vobs0}) and (\ref{fringe}) provide
a clear-cut argument to understand why the fringe shifts were coming out
 much smaller than classically 
expected: they are proportional to the squared Earth's velocity 
    through the Fresnel's drag coefficient of the dielectric medium used 
    in the interferometer. Thus, there should be no surprise that the
 `observable' velocity is much smaller than the `kinematical' velocity. 

Also, the trend predicted by 
eqs.(\ref{vobs0}) and (\ref{fringe}) is such to reproduce correctly
the experimental results. In fact,
 the observable velocity, and thus the anisotropy, becomes smaller
and smaller when ${\cal N}_{\rm medium}$ approaches unity and vanishes
identically
in the limit ${\cal N}_{\rm medium}\to 1$. This is consistent with the analysis
of the experiments performed by 
Kennedy, Illingworth and Joos vs. those of
Michelson-Morley, Morley-Miller and Miller.
We note that a qualitatively 
similar suppression effect had already been discovered by
Cahill and Kitto \cite{cahill} by following a different approach. 

\section{Interpretation of the classical ether-drift experiments}

Now, if upon operation of the interferometer
there are fringe shifts  and if their magnitude,
 observed with different dielectric media and within the
experimental errors, points
consistently to a unique value of the kinematical Earth's velocity, there is
experimental evidence for
the existence of a preferred frame $\Sigma \neq S'$. In practice, to
 ${\cal O}({{v^2_{\rm earth} }\over{c^2}} )$, this can be decided by
re-analyzing the experiments in terms of the effective parameter
 $\epsilon = {{v^2_{\rm earth} }\over{u^2}} k_{\rm medium}$. The
conclusion of Cahill and Kitto \cite{cahill} 
is that the classical experiments are consistent with the value
 $v_{\rm earth}\sim 365$ km/s obtained from the dipole fit to the COBE data
\cite{cobe} for the anisotropy
of the cosmic background radiation. 

However, in
our expression eq.(\ref{vobs0}) determining the fringe shifts there is a difference
of a factor $\sqrt{3}$ with respect to their result
$v_{\rm obs}=v \sqrt { k_{\rm medium} }$. Therefore, using
eqs.(\ref{vobs0}) and (\ref{vobs}), for ${\cal N}_{\rm air} \sim 1.00029$, 
 the relevant Earth's velocity (in the plane of the interferometer)
 is {\it not} $v_{\rm earth}\sim 365$ km/s but rather
\BE
\label{vearth}
                  v_{\rm earth} \sim 204 \pm 36 ~{\rm km/s}
\EE
This value provides a definite range of velocities that can be
used in the analysis of the other experiments. 

To this end, let us compare with the experiment performed by Michelson, 
Pease and Pearson \cite{mpp}. These other authors in 1929, 
using their own interferometer, again at Mt. Wilson, declared that 
their ``precautions taken to eliminate effects of 
temperature and flexure disturbances were effective''. Therefore, their statement that the
fringe shift, 
as derived from ``...the displacements observed at maximum and minimum at 
sidereal times...'', was definitely smaller than ``...one-fifteenth of
that expected on the 
supposition of an effect due to a motion of the Solar System of three 
hundred kilometres per second'', can be taken as an indirect 
confirmation of our eq.(\ref{vearth}). Indeed,
although the ``one-fifteenth'' was actually 
a ``one-fiftieth'' (see page 240 of ref.\cite{miller}), 
their fringe shifts were certainly non negligible. This is easily understood 
since, for an in-air-operating interferometer, 
the fringe shift $(\Delta\lambda)_{\rm class}(300)$, expected on the base of 
classical physics
for an Earth's velocity of 300 km/s, is about 500 times
bigger than the corresponding relativistic one
\BE
(\Delta\lambda)_{\rm rel}(300)\equiv 3 k_{\rm air}
~ (\Delta\lambda)_{\rm class}(300)
\EE
computed using Lorentz transformations 
(compare with eq.(\ref{fringe}) for  
$k_{\rm air}\sim {\cal N}^2_{\rm air} -1 \sim 0.00058$). 
 Therefore, the Michelson-Pease-Pearson upper bound
\BE
(\Delta\lambda)_{\rm obs}< 0.02~
 (\Delta\lambda)_{\rm class} (300)
\EE
is actually equivalent to
\BE
(\Delta\lambda)_{\rm obs}< 24 ~
 (\Delta\lambda)_{\rm rel} (204)
\EE
As such, it poses no strong restrictions and is entirely 
consistent with those typical low observable velocities reported in 
eq.(\ref{vobs}).

A similar agreement is obtained when comparing with the Illingworth's data
\cite{illing} as recently re-analyzed by M\'unera \cite{munera}. 
In this case, using eq.(\ref{vobs0}), 
the observable velocity $v_{\rm obs}=3.1 \pm 1.0$ km/s \cite{munera}
(errors at the 95$\%$ C.L.)  
and the value ${\cal N}_{\rm helium}-1 \sim 3.6\cdot 10^{-5}$, 
one deduces $v_{\rm earth}=213 \pm 36$ km/s
(errors at the 68$\%$ C.L.) in very good agreement with our
eq.(\ref{vearth}). 

The same conclusion applies to the Joos experiment 
\cite{joos}. Although we don't know the exact 
value of ${\cal N}_{\rm vacuum}$ for the Joos experiment, 
it is clear that his result, $v_{\rm obs}<$ 1.5 km/s, represents
the natural type of upper bound
in this case. As an example, for $v_{\rm earth}\sim 204$ km/s, one
obtains $v_{\rm obs}\sim 1.5$ km/s for 
${\cal N}_{\rm vacuum}-1= 9\cdot 10^{-6}$ and 
$v_{\rm obs}\sim 0.5$ km/s for 
${\cal N}_{\rm vacuum}- 1=1\cdot 10^{-6}$. 
In this sense, the effect of using Lorentz 
transformations is most 
dramatic for the Joos experiment when comparing with
the classical expectation for an Earth's velocity
of 30 km/s. Although the relevant Earth's velocity 
can be as large as $204$ km/s, 
the fringe shifts, rather than being $(204/30)^2\sim 50$ times {\it bigger} than the 
classical prediction, are  
$\sim (30/1.5)^2= 400$ times {\it smaller}. 

Notice that, using our eq.(\ref{vobs0}), 
the kinematical Earth's velocity obtained from the {\it absolute magnitude}
of the fringe shifts becomes consistent with that 
needed by Miller to understand the {\it variations} 
of the ether-drift effect in different epochs of the year \cite{miller}.
In fact, the typical
daily values, in the plane of the interferometer, had to lie in the range
195 km/s$ \le v_{\rm earth} \le 211$ {\rm km/s} 
(see table V of ref.\cite{miller}).  
Such a consistency, on one hand, 
increases the body of experimental evidence for a preferred frame, and
on the other hand, signals the internal consistency of Miller's analysis. 

We are aware that our conclusion goes against the widely spread belief, 
originating from the paper of Shankland et {\it al.} ref.\cite{shankland}, 
that Miller's results were actually due to statistical fluctuation and/or 
local temperature conditions. 
To a closer look, however, the argument of Shankland et {\it al.} is not so 
solid as it appears by reading the Abstract of their paper. In fact, 
within the paper these authors say that ``...there can be 
little doubt that statistical fluctuations alone  cannot account for the
periodic fringe shifts observed by Miller'' 
(see page 171 of ref.\cite{shankland}). In fact, although ``...there is 
obviously considerable scatter in the data at each azimuth position,...the
average values...show a marked second harmonic effect''
(see page 171 of ref.\cite{shankland}). In any case, interpreting the observed
effects on the base of the local temperature conditions is certainly not 
the only explanation since ``...we must admit that a direct and general 
quantitative correlation between amplitude and phase of the observed 
second harmonic on the one hand and the thermal conditions in the observation
hut on the other hand could not be established'' 
(see page 175 of ref.\cite{shankland}). This rather
unsatisfactory explanation of the observed effects should be compared 
with the previously mentioned
excellent agreement that was instead obtained by Miller
once the final parameters for the Earth's velocity were plugged in the 
theoretical predictions (see figs.26 and 27 of ref.\cite{miller}).

The most surprising thing, however, is that Shankland et {\it al.} did not realize
that Miller's average value 
${A}_2=0.044 \pm 0.005$, obtained after their own critical re-analysis
of his observations at Mt.Wilson, when compared to the expected classical 
value $A_2=0.56$ for his interferometer, was
giving precisely the same $v_{\rm obs}\sim 8.4 \pm 0.5$ km/s 
obtained from the Miller's re-analysis of the Michelson-Morley 
experiment in Cleveland. Conceivably, 
their emphasis on the role of the temperature effects
in the Miller's data would have been re-considered
whenever they had realized the perfect identity
of two determinations obtained 
in completely different experimental conditions.

\section{Comparison with present-day experiments}

Let us finally consider those present-day, 
 `high vacuum' Michelson-Morley experiments of the type first performed by 
Brillet and Hall \cite{brillet} and more recently by 
M\"uller et {\it al.} \cite{muller}. 
In these experiments, the test of the isotropy 
of the speed of light does not consist in the observation of the interference
fringes as in the classical experiments. 
Rather, one looks for the difference
$\Delta \nu$
in the relative frequencies of two cavity-stabilized lasers upon local
rotations of
the apparatus \cite{brillet} or under the Earth's rotation \cite{muller}
on the base of the relation
\BE
\label{nutheta}
  \nu (\theta)= {{ \bar{u}'(\theta) n}\over{2L}}
\EE 
Here $\bar{u}'(\theta)$ is the two-way speed of light within the cavity, 
$n$ is the integer number fixing the cavity mode and $L$ the length
of the cavity as measured in $S'$. Again, as stressed in connection with 
eq.(\ref{fringe}), due to the isotropy of space the cavity length 
is taken to be independent of the cavity orientation.

The present experimental value for the anisotropy of the two-way speed of light
in the vacuum, as determined by M\"uller et {\it al.}\cite{muller}, 
\BE
\label{experiment}
{{\Delta \nu_\theta}\over{\nu}}=
        ({{ \Delta \bar{c}_\theta }\over{c}})_{\rm exp}= (2.6 \pm 1.7) \cdot 10^{-15}
\EE
can be interpreted
within the framework of our eq.(\ref{twoway}) where
\BE
        ({{ \Delta \bar{c}_\theta }\over{c}})_{\rm theor} \sim 
|B_{\rm vacuum}| {{v^2_{\rm earth} }\over{c^2}} 
\EE
Now, in a perfect vacuum by definition 
${\cal N}_{\rm vacuum}=1$ so that
$B_{\rm vacuum}$ and  $v_{\rm obs}$ vanish. 
However, one can explore 
\cite{pagano} the possibility that, even in this case,
 a very small anisotropy might be due to a refractive index 
${\cal N}_{\rm vacuum}$ that differs from unity by an infinitesimal
amount. In this case, the natural candidate to explain a value
${\cal N}_{\rm vacuum} \neq 1$
is gravity. In fact, by using
the Equivalence Principle, a freely falling frame $S'$ will locally 
measure the same speed of light as in an inertial frame in the absence of
any gravitational effect. However, if $S'$ carries on board an heavy 
object this is no longer true. For an observer placed on the Earth, 
this amounts to insert
the Earth's gravitational potential in the  weak-field isotropic
approximation to the line element of
 General Relativity \cite{weinberg}
\BE
ds^2= (1+ 2\varphi) dt^2 - (1-2\varphi)(dx^2 +dy^2 +dz^2)
\EE
so that one obtains a refractive index for
light propagation 
\BE
\label{nphi}
            {\cal N}_{\rm vacuum}\sim  1- 2\varphi
\EE
This represents the `vacuum analogue' of 
${\cal N}_{\rm air}$, ${\cal N}_{\rm helium}$,...so that from
\BE 
     \varphi =- {{G_N M_{\rm earth}}\over{c^2 R_{\rm earth} }} \sim
-0.7\cdot 10^{-9}
\EE
and using eq.(\ref{BTH}) one predicts
\BE
\label{theor}
                 B_{\rm vacuum} \sim -4.2 \cdot 10^{-9}
\EE
Adopting the range
of Earth's velocity (in the plane
of the interferometer) given in eq.(\ref{vearth}) this leads to predict
an observable anisotropy of the two-way speed of light 
in the vacuum eq.(\ref{twoway}) 
\BE
\label{theory}
        ({{ \Delta \bar{c}_\theta }\over{c}})_{\rm theor} \sim 
|B_{\rm vacuum}| {{v^2_{\rm earth} }\over{c^2}} \sim (1.9\pm 0.7)\cdot 10^{-15}
\EE
consistently with the experimental value in eq.(\ref{experiment}). 

Clearly, in this framework, trying to rule out the existence of
a preferred frame through the experimental 
determination of ${{ \Delta \bar{c}_\theta }\over{c}}$ in a high
vacuum is not the most convenient strategy 
due to the vanishingly small value of
$B_{\rm vacuum}$. In other words, even with years of data taking \cite{muller}, 
it is not easy to rule out the theoretical prediction in eq.(\ref{theory})
starting from the present experimental value eq.(\ref{experiment}). 

For this reason, a more efficient search 
might be performed in dielectric gaseous media where, if there is a preferred
frame, the frequency of the signal should be much larger. 
As a check, we have compared with the only available results
obtained by Jaseja et. al \cite{jaseja} in 1963 when
looking at the relative frequency shifts of two orthogonal 
He-Ne masers placed on a rotating platform. 
As we shall show in the following, their data are consistent with the same type
of conclusion obtained from the classical experiments: an ether-drift effect 
determined by an Earth's velocity as in eq.(\ref{vearth}). 

To use the experimental results reported by Jaseja et {\it al.}\cite{jaseja}
one has to subtract preliminarly a large overall systematic 
effect that was present in their data and interpreted by the authors as
probably due to magnetostriction
in the Invar spacers induced by the Earth's magnetic field. 
As suggested by the same authors, this spurious
effect, that was only affecting the normalization of the experimental
$\Delta \nu$, can be subtracted looking at the variations of the data
at different hours of the day. The data for $\Delta\nu$,
in fact, in spite of their rather large errors,
exhibit a characteristic modulation (see fig.3 of ref.\cite{jaseja})
with a maximum at about 7:30 a.m. and a minimum at about
9:00 a.m.. To estimate the size of the time modulation, one can
follow two different strategies: a) just consider the two data corresponding to
the maximal and minimal values
$\Delta \nu_{\rm exp} (7:30~{\rm a.m.})\sim 276 \pm 5$ kHz, 
$\Delta \nu_{\rm exp} (9:00~{\rm a.m.})\sim 267 \pm 4$ kHz
and the difference
\BE
\label{resulta}
 \delta_a(\Delta\nu)\equiv  
\Delta \nu_{\rm exp} (7:30~{\rm a.m.}) -
\Delta \nu_{\rm exp} (9:00~{\rm a.m.}) \sim (9\pm 6)~ {\rm kHz}
\EE
or b), following Jaseja et {\it al.}, group the data in two bins of six
by defining average values, say
$\langle\Delta \nu\rangle_{\rm exp} (7:30~{\rm a.m.})$ and
$\langle\Delta \nu\rangle_{\rm exp} (9:00~{\rm a.m.})$, thus obtaining
\BE
\label{resultb}
 \delta_b(\Delta\nu)\equiv 
\langle\Delta \nu\rangle_{\rm exp} (7:30~{\rm a.m.})-
\langle\Delta \nu\rangle_{\rm exp} (9:00~{\rm a.m.})
\sim (1.6\pm 1.2)~ {\rm kHz}
\EE
Our theoretical starting point to understand the above 
(rather loose) determinations
is the formula for the frequency shift of the two
masers at an angle $\theta$ with the direction of the ether-drift 
\BE
\label{prediction}
              {{\Delta \nu (\theta) }\over{\nu}}= 
{{\bar{u}'(\pi/2 +\theta)- \bar{u}' (\theta)} \over{u}}=
|B_{\rm He-Ne}| {{v^2_{\rm earth} }\over{c^2}} \cos(2\theta)
\EE
where, taking into account the values
${\cal N}_{\rm helium}\sim 1.000036$, ${\cal N}_{\rm neon}\sim 1.000067$, 
${\cal N}_{\rm He-Ne}\sim 1.00004$ and eq.(\ref{BTH}) we shall use
$|B_{\rm He-Ne}|\sim 1.2\cdot 10^{-4}$. 

Further, using the value of the frequency of ref.\cite{jaseja} 
$\nu\sim 3\cdot 10^{14}$ Hz and our
standard value eq.(\ref{vearth})
for the Earth's velocity in the plane of the interferometer 
$v_{\rm earth} \sim 200$ km/s, eq.(\ref{prediction}) leads to
the reference value for the amplitude of the signal 
\BE
\label{reference}
(\Delta \nu)_{\rm ref}= \nu
|B_{\rm He-Ne}| {{(200~ {\rm km/s})^2 }\over{c^2}} \sim 16 ~{\rm kHz}
\EE
and to its time modulation
\BE
\label{prediction2}
 \delta(\Delta\nu)_{\rm theor}
 \sim 16~{\rm kHz} ~{{\delta v^2}\over{v^2}}
\EE
where
\BE
\label{deltav}
{{\delta v^2}\over{v^2}}\equiv 
 {{v^2_{\rm earth} (7:30~{\rm a.m.}) -v^2_{\rm earth} (9:00~{\rm a.m.}) }
\over{(200~{\rm km/s})^2}}
\EE
To evaluate the above ratio of velocities, let us first 
compare the modulation of $\Delta\nu$ seen in fig.3 of ref.\cite{jaseja} 
with that of $v_{\rm obs}$ in
fig.27 of ref.\cite{miller} (data plotted as a function of civil time as
in ref.\cite{jaseja}) restricting to the 
Miller's data of February, the period of the year that is
closer to the date of January 20th when Jaseja et {\it al.} performed their
experiment. Further, 
the different location of the two laboratories (Mt.Wilson
and Boston) can be taken into account with a shift of about three hours so that
Miller's interval 3:00 a.m.$-$9:00 a.m. is made to correspond
to the range 6:00 a.m.$-$12:00 a.m. of
Jaseja et {\it al.}. If this is done, 
although one does not expect an exact correspondence
due to the difference between the 
two epochs of the year, the two characteristic trends are surprisingly
close. 

Thus we shall try to use the Miller's data 
for a rough evaluation of the ratio reported in eq.(\ref{deltav}) after
rescaling from $v_{\rm obs}$ 
to $v_{\rm earth}$ through eq.(\ref{vobs0}) (for the Miller's interferometer
that was operating in air).
Following for the Miller's data the same procedure used to obtain
eq.(\ref{resulta}) ({\it i.e.} just restricting to the difference between maximal 
and minimal values) we  obtain 
a maximal observable velocity 
$(v_{\rm obs})^{\rm max}\sim 9.4$ km/s, that corresponds to a value
$(v_{\rm earth})^{\rm max}\sim 225$ km/s,  and
a minimal observable velocity 
$(v_{\rm obs})^{\rm min}\sim 7.5$ km/s, that corresponds to a value
$(v_{\rm earth})^{\rm min}\sim 180$ km/s. These velocities, when replaced in
eq.(\ref{deltav}), produce a value
${{(\delta v^2)_a}\over{v^2}} \sim 0.46$ that, 
when used in eq.(\ref{prediction2}), leads 
to a theoretical prediction $\delta(\Delta\nu)_{\rm theor}\sim 7.3$ kHz, 
 well consistent with the experimental result in
eq.(\ref{resulta}). On the other hand, averaging the Miller's data slightly
to the left and to the right of the minimum, we get the
smaller value ${{(\delta v^2)_b}\over{v^2}} \sim 0.12$
that, when replaced in eq.(\ref{prediction2}), leads 
to $\delta(\Delta\nu)_{\rm theor}\sim 1.8$ kHz
 consistently with eq.(\ref{resultb}). 
Of course, for a really significative test,
one needs more precise data. However, with the present data 
eqs.(\ref{resulta}) and (\ref{resultb}), and
 in spite of our 
crude approximations, the order of magnitude of the effect is correctly
reproduced. 

This suggests, once more \cite{pagano}, 
to perform a new class of ether-drift experiments in 
dielectric gaseous media. For instance,
using stabilizing cavities as in Refs.\cite{brillet,muller}, one could
replace the high vacuum 
in the Fabry-Perot with air. In this case, where
$|B_{\rm vacuum}|\sim 4 \cdot 10^{-9}$ would be replaced by
$|B_{\rm air}|\sim 9\cdot 10^{-4}$, there should be an increase by
five orders of magnitude in the typical value of $\Delta \nu$ with respect to
refs.\cite{brillet,muller}.

\section{Summary and outlook}

In this paper we have re-considered
the possible existence of a preferred reference 
frame through an analysis of the classical and modern ether-drift experiments. 
Our re-analysis started with the original data obtained
by Michelson and Morley \cite{mm} in each session of
their experiment. Contrary to the generally accepted ideas, 
 but in agreement with the point of view 
expressed by Hicks in 1902 \cite{hicks}, Miller in 1933 \cite{miller}
 and M\'unera in 1998 \cite{munera}, 
the results of that experiment
should {\it not} be considered null. The even combinations of 
fringe shifts 
  ${{\Delta \lambda(\theta) + \Delta \lambda(\pi+\theta)}\over{2\lambda}} $, 
although smaller than
the classical prediction corresponding
to the orbital motion of the Earth, 
 exhibit the characteristic second-harmonic behaviour (see fig.2)
expected for an ether-drift effect. The average amplitude of the second-harmonic
component 
${A}_2\sim 0.016 \pm 0.002$ (see table 2), when normalized to the expected
classical value
${{D}\over{\lambda}} ~{{(30 {\rm km/s})^2 }\over{c^2}}\sim 0.2$ for the
Michelson-Morley interferometer, corresponds to an
average observed velocity 
$v_{\rm obs} \sim 8.4 \pm 0.5$ km/s.
 
As emphasized at the end
of Sect.2 and at the end of Sect.4, 
this average value of $v_{\rm obs}$ for the Michelson-Morley experiment
is {\it exactly} the
same average daily value that was obtained by Miller in his 1925-1926 
observations at Mt.Wilson. This can easily be checked, after the critical 
re-analysis of Shankland et {\it al.}, by comparing
 Miller's average daily value ${A}_2\sim 0.044 \pm 0.005$ 
(see page 170 of ref.\cite{shankland}) with the expected classical value
${{D}\over{\lambda}} ~{{(30 {\rm km/s})^2 }\over{c^2}}\sim 0.56$ for the
Miller's interferometer. 

 Our conclusion is further confirmed by the independent analysis 
of the available Miller's data performed by M\'unera \cite{munera}
which provides the value $v_{\rm obs} \sim 8.2 \pm 1.4$ km/s.
By including the other determinations 
obtained by Morley and Miller in the period
1902-1905 (see fig.4 of ref.\cite{miller}), the results of 
these three main classical ether-drift experiments can be summarized in the
value $v_{\rm obs} \sim 8.5 \pm 1.5$ km/s.

Therefore, once different ether-drift experiments give consistent values
of the Earth's observable
velocity, it becomes natural to explore the existence of a preferred frame, 
within the context of Lorentzian Relativity, and
use Lorentz transformations 
to extract the real kinematical velocity corresponding to this
$v_{\rm obs}$. In this case, using eq.(\ref{vobs0}), 
we find a value (in the plane of the interferometer)
$v_{\rm earth} \sim 204 \pm 36 $ km/s 
in agreement with the range 195 km/s $\leq v_{\rm obs} \leq $ 211 km/s
 needed by Miller to describe the 
variations of the ether-drift effect in different epochs of the year
(see table V of ref.\cite{miller}). 

At the same time, using Lorentz transformations, the same range
of $v_{\rm earth}$ corresponds to an
effective $v_{\rm obs}\sim 3$ km/s 
for the Kennedy's \cite{kennedy0} and Illingworth \cite{illing} 
experiments (performed in helium) or
$\sim 1$ km/s for the Joos experiment \cite{joos} (performed in an evacuated
housing) consistently with the experimental results. 

Additional checks of this theoretical framework are obtained by comparing
with the experimental data for the relative frequency shift $\Delta\nu$
which is measured in
the present-day experiments with cavity-stabilized lasers, upon local rotation
of the apparatus or under the Earth's rotation. In this case,
our basic relation is
\BE
{{\Delta\nu}\over{\nu}} \sim |B_{\rm medium}| {{v^2_{\rm earth}}\over{c^2}}
\EE
where $B_{\rm medium}\sim -3 ({\cal N}_{\rm medium}-1) $, 
${\cal N}_{\rm medium}$ being the refractive index of the gaseous dielectric
medium that fills the cavities. For a very high vacuum, using the 
prediction of General Relativity for an apparatus placed on the 
Earth's surface, $|B_{\rm vacuum}| \sim 4 \cdot 10^{-9}$, and the range
of kinematical Earth's velocity $v_{\rm earth}\sim 204\pm 36$ km/s
suggested by the classical ether-drift experiments, we predict
$({{\Delta\nu}\over{\nu}})_{\rm theor} \sim (1.9 \pm 0.7)\cdot 10^{-15}$, consistently
with the experimental result 
$({{\Delta\nu}\over{\nu}})_{\rm exp} = (2.6 \pm 1.7)\cdot 10^{-15}$ obtained
in ref.\cite{muller}.  

For He-Ne masers, the same range of Earth's velocities leads to predict
a typical value
$\Delta\nu\sim 16$ kHz, for which
${{\Delta\nu}\over{\nu}} \sim 5\cdot 10^{-11}$,  with a characteristic
modulation of a few kHz 
in the period of the year and for the hours of the day when
Jaseja et {\it al.}\cite{jaseja} performed their experiment. This prediction is
consistent with their data, although the rather large experimental 
errors require further experimental checks. To this end, an efficient
search for a preferred frame requires a modified experimental set-up where
the high vacuum adopted in the resonating cavities is replaced by air. 
In this case, where the anisotropy 
parameter $|B_{\rm vacuum}|\sim 4 \cdot 10^{-9}$ would be 
replaced by $|B_{\rm air}|\sim 9\cdot 10^{-4}$, there should be an increase of
{\it five orders of magnitude} in the typical 
value of $\Delta \nu$ with respect to
Refs.\cite{brillet,muller}. If such enhancement is not observed, rather than
waiting for years, the 
existence of a preferred frame will be definitely ruled out in a few days of
data taking.

\vfill
\eject
\begin{table}[t]
\centering{\begin{tabular}{c||c|c|c|c|c|c}  
 i      &  July 8 (n.) & July 9 (n.) & July 11 (n.)     
       &  July 8 (e.) & July 9 (e.) & July 12 (e.)     \\
1      & -0.001 & +0.018 &+0.015& -0.016& +0.007& +0.034 \\
2      & +0.024 & -0.004 &-0.035& +0.008& -0.015& +0.042 \\    
3      & +0.053 & -0.004 &-0.039& -0.010& +0.006& +0.045 \\   
4      & +0.015 & -0.003 &-0.067& +0.070& +0.004& +0.025 \\   
5      & -0.036 & -0.031 &-0.043& +0.041& +0.027& -0.004 \\  
6      & -0.007 & -0.020 &-0.015& +0.055& +0.015& -0.014 \\ 
7      & +0.024 & -0.025 &-0.001& +0.057& -0.022& +0.005 \\   
8      & +0.026 & -0.021 &+0.027& +0.029& -0.036& -0.013 \\    
9      & -0.021 & -0.049 &+0.001& -0.005& -0.033& -0.030 \\   
10     & -0.022 & -0.032 &-0.011& +0.023& +0.001& -0.066 \\    
11     & -0.031 & +0.001 &-0.005& +0.005& -0.008& -0.093 \\ 
12     & -0.005 & +0.012 &+0.011& -0.030& -0.014& -0.059 \\     
13     & -0.024 & +0.041 &+0.047& -0.034& -0.007& -0.040 \\       
14     & -0.017 & +0.042 &+0.053& -0.052& +0.015& +0.038 \\      
15     & -0.002 & +0.070 &+0.037& -0.084& +0.026& +0.057 \\      
16     & +0.022 & -0.005 &+0.005& -0.062& +0.024& +0.041 \\      
17     & -0.001 & +0.018 &+0.015& -0.016& +0.007& +0.034 \\     
\end{tabular}
\vskip 60 pt
\caption{\rm
We report the fringe shifts ${{\Delta \lambda(i)}\over{\lambda}}$
for all noon (n.) and evening (e.) sessions of the Michelson-Morley experiment.}
\label{tab:1}}
\end{table}

\bigskip
\bigskip
\bigskip
\bigskip

\vfill
.
\eject
\newpage


\begin{table}[t]
\centering{\begin{tabular}{c|c}  
SESSION       &       ${A}_2$   \\
July 8  (noon) & $0.010 \pm 0.005$  \\ 
July 9  (noon) & $0.015 \pm 0.005$   \\
July 11 (noon) & $0.025 \pm 0.005$    \\
July 8  (evening) & $0.014 \pm 0.005$  \\
July 9  (evening) &$0.011 \pm 0.005$   \\
July 12 (evening) & $0.018 \pm 0.005$  \\ 
\end{tabular}
\vskip 20 pt
\caption{\rm
We report the amplitude of the second-harmonic component
${A}_2$ obtained from the fit eq.(\ref{fourier}) 
to the various samples of data.}
\label{tab:2}}
\end{table}
\newpage

\begin{figure}[ht]
\psfig{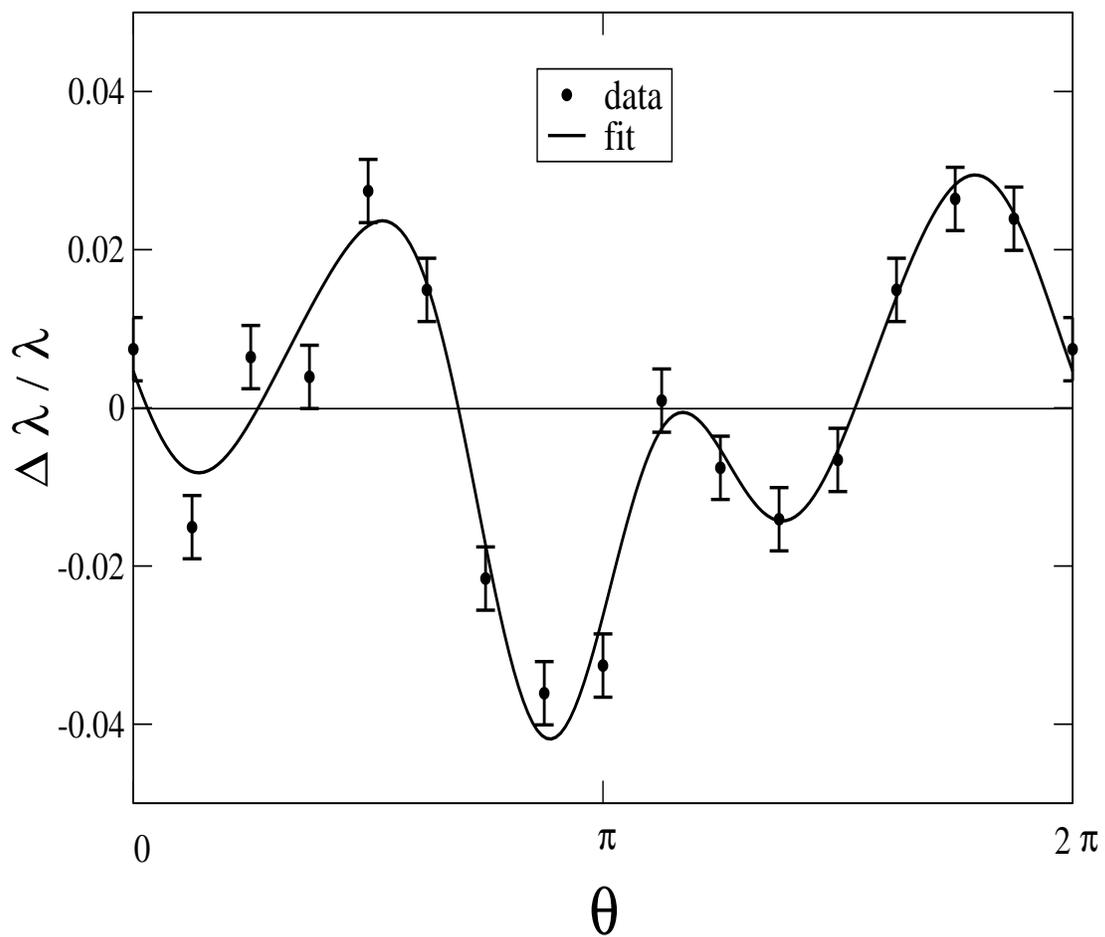}
\caption{
We show a typical fit eq.(4) to the Michelson-Morley data reported in Table 1. 
}
\end{figure}

\begin{figure}[ht]
\psfig{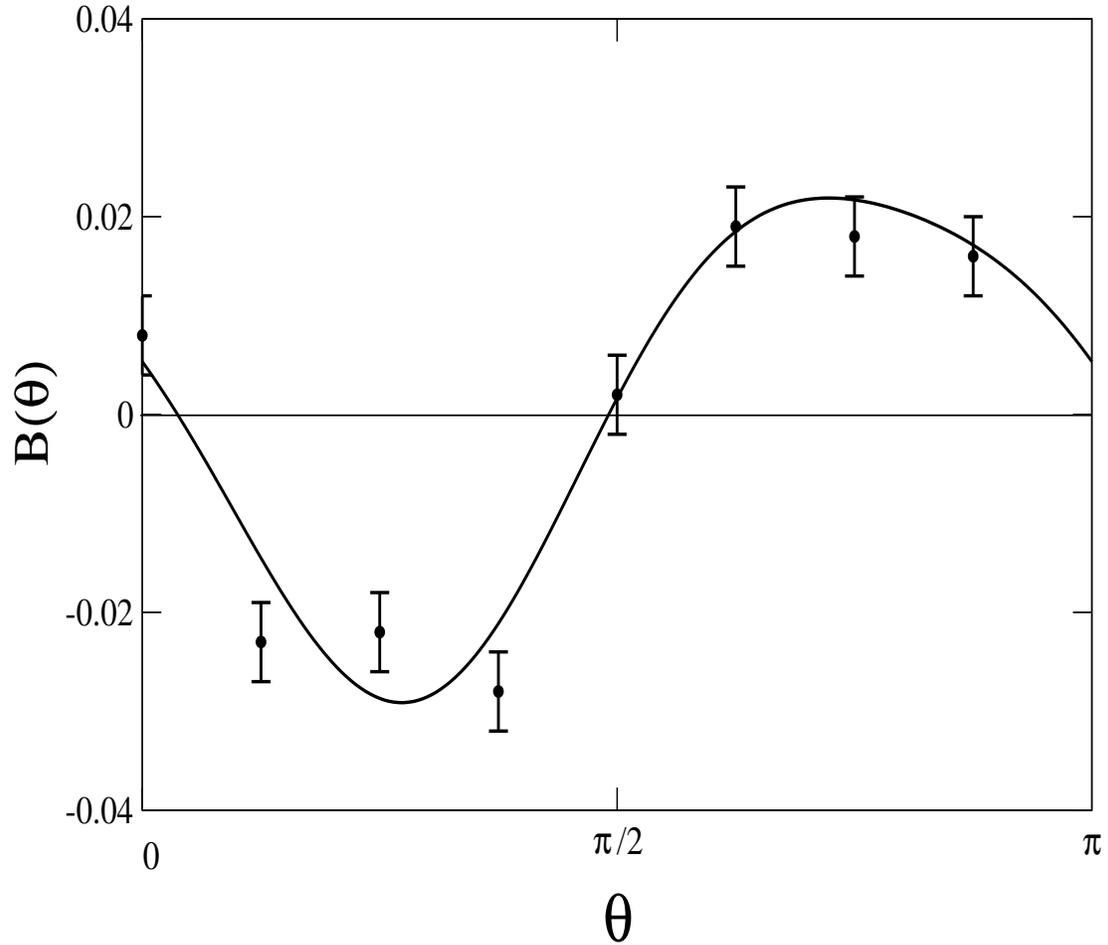}
\caption{
A typical fit eq.(4) to the even combination of fringe shifts
  $B(\theta)={{\Delta \lambda(\theta) + \Delta \lambda(\pi+\theta)}\over{2\lambda}} $
obtained from the
data reported in Table 1. The fitted amplitudes 
are ${A}_2=0.025 \pm 0.004$ and 
${A}_4=0.004 \pm 0.004$. }
\end{figure}

\end{document}